\documentclass[%
 reprint,
 amsmath,amssymb,
 aps,
onecolumn
]{revtex4-2}

\usepackage{crop}%
\usepackage{amsmath,amssymb,amsfonts,upref,endnotes}
\usepackage{amsthm}
\usepackage{marginnote}
\usepackage{soul}
\RequirePackage{graphicx}
\usepackage[figuresright]{rotating}
\usepackage{color}
\usepackage{graphicx}
\usepackage{dcolumn}
\usepackage{bm}
\usepackage{hyperref}

\begin{document}

\preprint{}

\title{Insensitive nonreciprocal edge breathers}

\author{Bertin Many Manda}
 \email{bmany@tauex.tau.ac.il}
 \affiliation{%
 School of Mechanical Engineering, Tel Aviv University, Tel Aviv 69978, Israel
 }
\affiliation{%
  Laboratoire d’Acoustique de l’Universit\'e du Mans (LAUM), UMR 6613, Institut d'Acoustique - Graduate School (IA-GS), CNRS, Le Mans Universit\'e, Av. Olivier Messiaen, 72085 Le Mans, France 
}%
\author{Vassos Achilleos}%
 \email{achilleos.vassos@univ-lemans.fr}
\affiliation{%
  Laboratoire d’Acoustique de l’Universit\'e du Mans (LAUM), UMR 6613, Institut d'Acoustique - Graduate School (IA-GS), CNRS, Le Mans Universit\'e, Av. Olivier Messiaen, 72085 Le Mans, France 
}%




\date{\today}

\begin{abstract}
We uncover subtle and previously unexplored phenomena arising from the interplay of nonlinearity and nonreciprocity in topological mechanical metamaterials.
We study a nonreciprocal topological Klein–Gordon chain of asymmetrically coupled nonlinear oscillators, which serves as a minimal mass-spring model capturing the features of several active nonreciprocal metamaterials across mechanical, electronic, and acoustic platforms.
We demonstrate that continuous families of nonreciprocal edge breathers (NEBs), namely boundary-localized, time-periodic waves, emerge from the linear edge mode as its amplitude increases.
Remarkably, despite the absence of chiral or sublattice symmetries, we identify insensitive NEBs whose nonlinear frequency remains fixed to that of the linear edge mode with increasing nonlinearity.
Our analysis reveals that the mechanism underlying this insensitivity stems from a competition between mode nonorthogonality and nonlinear interactions, yielding an exponential decay of the NEB nonlinear frequency shift with system size.
Crucially, these insensitive NEBs also persist in the strongly nonlinear regime.
Our work establishes a novel pathway toward realizing robust nonlinear topological waves in mechanical metamaterials without relying on symmetry-protected nonlinearities.
\end{abstract}

\maketitle


\section{\label{sec:introduction} Introduction}
Over the past several years, topological and nonreciprocal systems have emerged as fertile platforms for realizing unconventional wave phenomena~\cite{HK2010,KSUS2019,MXC2019,OPAGHLRSSZC2019,AGU2020,WC2023,LTLL2023,NYKA2023,ZZCLC2023,SBPM2024}. 
On one hand, topological mechanisms give rise to protected edge modes that remain robust against disorder and defects, as demonstrated in theory and experiment in photonic~\cite{LJS2014,OPAGHLRSSZC2019,OPAGHLRSSZC2019,SLCK2020,SB2021,ZZCLC2023}, atomic~\cite{YLXF2018,EWMHS2023,YSWSSCHLQWDHXCDGY2025}, phononic~\cite{H2016,SH2016,YRWYCY2018,MXC2019,XSHMY2020,HCH2021,SBPM2022,ZZCLC2023,LMLYTD2024,SBPM2024} and electronic~\cite{HK2010,QZ2011,E2018a,SLCK2020} systems.
On the other hand, nonreciprocal couplings introduce a directional bias that leads to the non-Hermitian skin effect (NHSE), whereby all bulk modes accumulate at one boundary of a finite lattice with open boundary conditions (OBCs)~\cite{KEBB2018,LT2019,YW2018,OKSS2020,LTLL2023,OS2023,WC2023,ZZCLC2023}, see also Refs.~\cite{E2022c,L2022}. 
The NHSE has demonstrated to be universal, with experimental observations across the same broad range of platforms as topological systems~\cite{BLLC2019,GBVC2020,WKHHSGTS2020,LSMZYWJJZ2021,ZYGGCYCXLJYSCZ2021,ASPMCACBOPG2023,MAPPA2023,JMAFS2025,PVRRVF2025,LDHKL2025,SJL2025,WWLQZLLL2025}. 
In this context, the interplay between the topological edge mode and the NHSE has shown to be capable of reshaping its wave function, enabling localization at either boundaries or even extended protected modes~\cite{ZTLG2021,WWM2022,LCHWLHDL2024}, stretching far beyond the possibilities of reciprocal systems.

On the other hand, nonlinearity is an intrinsic feature of many physical systems, including optical~\cite{LSCASS2008}, mechanical~\cite{BRBT2021}, acoustic~\cite{IRATDF2023}, atomic~\cite{TBLG2020}, and electronic~\cite{SJL2025} platforms. This has motivated extensive investigations into the interplay between topology and nonlinearity~\cite{PVLR2018,CT2019,DL2019,SLCK2020,E2022a,BLLFWX2024,GJPLDF2025}, as well as between nonreciprocity and nonlinearity~\cite{KMM2023,KN2025,SJL2025,WWLQZLLL2025}. 
These efforts have revealed a wealth of phenomena, including the nonlinear non-Hermitian skin effect~\cite{Y2021,JCZL2023,MCKA2024} and nonlinear topological edge solitons and breathers~\cite{SLCK2020,MS2021,J2023,BLLFWX2024}. 
More recently, it has been shown that the combined action of nonlinearity, nonreciprocity, and topology enables control over the shape and spatial localization of finite-amplitude edge states in photonic systems~\cite{CWZLN2025}. 
However, such tunability typically relies on nonlinearities that preserve the symmetries protecting the underlying topological edge modes, thereby preventing amplitude-dependent frequency shifts~\cite{CWZLN2025,GJPLDF2025}.

In general, nonlinearities tend to break the symmetries that protect topological edge modes. Nevertheless, even in the absence of such symmetries, the interplay between nonlinearity, nonreciprocity, and topology can give rise to unexpected phenomena. 
Notably, in nonreciprocal topological photonic systems with onsite Kerr nonlinearity, it was shown that insensitive nonreciprocal stationary solitons emerge from the zero-energy edge mode, with the nonlinear energy remaining identically zero as the amplitude increases, despite the absence of chiral or sublattice symmetries.
In light of recent experimental advances in the design and implementation of nonreciprocal mechanical metamaterials~\cite{WWM2022,VGBVTCC2025}, one may naturally ask whether such properties can be extended to these systems.
Indeed, the situation is qualitatively different in mechanical metamaterials.
In the linear limit, the edge modes occur at finite frequencies and display spatially oscillatory profiles~\cite{MXC2019,SBPM2022}.
As a result, the introduction of arbitrary nonlinearities naturally leads to time-periodic nonlinear excitations in the form of edge breathers~\cite{CXYKT2021}, which are fundamentally different from stationary states of photonic latices~\cite{CXYKT2021,MS2021}. 
This raises several fundamental and previously unexplored questions: Do nonreciprocal topological mechanical systems support nonreciprocal edge breathers (NEBs)? 
If so, can such NEBs exhibit spectral insensitivity to nonlinear perturbations breaking the symmetries protecting the edge modes?

In this work, we address these questions by investigating a topological Klein–Gordon (KG) chain of asymmetrically coupled nonlinear classical oscillators.
This model provides a minimal nonlinear mass–spring model for different physical realizations of active nonreciprocal metamaterials across mechanical~\cite{BLLC2019,WWM2022,VGBVTCC2025}, electronic~\cite{JMAFS2025}, and acoustic~\cite{ZYGGCYCXLD2021,MAPPA2023} platforms.
By combining multiple scale analysis and numerical continuation methods, we demonstrate that continuous families of NEBs emerge from the edge mode at any parameter values.
We then uncover a region of the parameter space of insensitive NEBs.
Our analysis reveal that this insensitivity originates from the competition between mode nonorthogonality rooted in the nonreciprocal couplings (non-Hermiticity) and their nonlinear interactions, resulting in a zero nonlinear frequency shift, thus  insensitive NEBs.
Remarkably, these insensitive NEBs persist deep into the strongly nonlinear regime.
We also numerically analyze their linear stability, addressing finite-size effects, among others.

The paper is structured as follows.
In Sec.~\ref{sec:system_skin_and_edge_modes}, we introduce the nonreciprocal topological Klein-Gordon chain and analyze its spectral properties in the linear limit.
Section~\ref{sec:skin_breathers} presents the families of NEBs obtained through numerical continuation, along with their Floquet stability and representative cases of their dynamics.
We then provide a theoretical explanation for the observed spectral insensitivity and discuss numerical results obtained beyond the perturbative limit.
Finally, in Sec.~\ref{sec:conclusion_breathers}, we summarize our findings.
Detailed analytical derivations of the NEB frequencies, wave functions, and associated sensitivities are provided in the appendices.

\section{\label{sec:system_skin_and_edge_modes}Model, skin and edge modes}
\begin{figure}[!t]
    \centering
    \includegraphics[width=0.7\columnwidth]{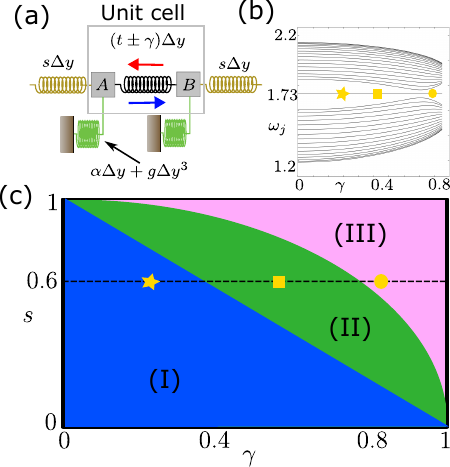}
    \caption{(a) Schematic of the unit cell of the nonreciprocal topological KG chain of nonlinear classical oscillators.
    The $s$ and $t\pm \gamma$ are the inter-and-intracell elastic stiffnesses, with $\gamma$ tuning the nonreciprocal strength.
    Here the elongation $\Delta y$ has the dimension of displacements.
    Further, the $\alpha$ and $g$ are the elastic and nonlinear stiffnessses of the onsite spring.
    (b) Dependence of the linear frequencies, $\omega_j$ against the nonreciprocal parameter, $\gamma$ numerically computed for a lattice of $N=33$ sites ($M=17$ cells) with $s=0.6$, see horizontal dashed line in (c).
    (c) Phase diagram of the nonreciprocal topological KG chain of classical oscillators.
    The phase diagram is obtained analyzing the localization and location of the right and left eigenvectors of the edge mode.
    The dashed line guides the eyes to $s=0.6$, with the star-, square- and dot-symbol indicating representative cases in regions (I), (II) and (III), with $\gamma=0.3$, $\gamma=0.6$ and $\gamma=0.825$ respectively, see text for details. 
    }
    \label{fig:chain_and_spectrum_lineear}
\end{figure}

We consider a mechanical variant of a nonreciprocal Su-Shrieffer-Heeger (SSH) lattice model with on-site (Kerr) nonlinearity~\cite{CZLC2022,MA2024}.
The equivalent mechanical unit cell, is formed of two classical oscillators with onsite nonlinear springs connected with asymmetric elastic couplings as shown in Fig.~\ref{fig:chain_and_spectrum_lineear}(a).
These units cells are then tight altogether with identical linear springs, forming a nonreciprocal topological Klein-Gordon (KG) chain of classical nonlinear oscillators.
It follows that the equations of motion are generated by
\begin{align}
    \begin{split} \label{eq:eq_motion}
        \ddot{y}_{A,m} &= (1+\gamma)\left(y_{B,m-1} - y_{A,n}\right) + s\left(y_{B,m} - y_{A,m}\right)  - \alpha_{A,m} y_{A,m} - gy_{A,m}^3,
    \end{split}
    \\[2ex] \nonumber
    \begin{split}
        \ddot{y}_{B,m} &= s\left(y_{A,m} - y_{B,m} \right)  - (1-\gamma)\left(y_{A,m+1} - y_{B,m}\right)  - \alpha_{B,n} y_{B,m} - gy_{B,m}^3,
    \end{split} &
\end{align}
where $y_{a,m}$ and $\dot{y}_{a,m}=dy_{a,m}/dt$ ($t$ is the time and $a=\{A,B\}$) are conjugate displacement and momentum, and the subscripts $\{A,m\}$ (resp. $\{B,m\}$) labels the first (resp. second) site within the unit cell of index $m$ [Fig.~\ref{fig:chain_and_spectrum_lineear}(a)].
The $1\pm \gamma$ and $s$ are the intracellular asymmetric and intercellular stiffnesses, $\alpha_{a,m}$ are the onsite elastic coefficients, and $g$ determines the onsite nonlinear strength.
Further, we consider a finite lattice of $M$ cells or equivalently $N$ oscillators with fixed-fixed boundary conditions.
Here, positive values of $g$ correspond to hardening-type nonlinearities, whereas negative ones lead to softening-type nonlinearities.
As mentioned in Sec.~\ref{sec:introduction} several experimental realizations which mimic such lattices, have already been implemented~\cite{WWM2022,VGBVTCC2025}.

The linearized limit of the system is obtained when the amplitude of the waves are small $\lvert y_{a,m}\rvert \rightarrow 0$.
Looking for solutions of the form $y_{a,m} = u_{a,m} e^{-i\omega t}$ leads to the right eigenvalue problem
\begin{equation}
    \omega^2 \vec{u} = D \vec{u},
    \label{eq:eigenvalue_problem}
\end{equation}
where $\omega$ is the oscillation frequency, and $\vec{u}$ is its associated wave function with
$\vec{u} = (u_{A,1}, u_{B,1}, u_{A,2}, u_{B,2}, \ldots, u_{A,M}, u_{B,M})$ or equivalently $\vec{u} = (u_1, u_2, u_3, \ldots, u_N)$.
Further,
\begin{equation}
D = \begin{pmatrix}
        2 + \alpha_0 & -s & 0 & \ldots & 0 & 0\\
       -s & -2 + \alpha_0 & -(1-\gamma) & \ldots & 0 & 0 \\
       0 & -(1+\gamma) &  2 + \alpha_0 & \ldots & 0 & 0 \\
       \vdots & \vdots & \vdots & \ddots & \vdots & \vdots \\
       0 & 0 & 0 & \ldots & 2 + \alpha_0 &  -(1-\gamma) \\
       0 & 0 & 0 & \ldots & -(1+\gamma) & 2 + \alpha_0 \\
   \end{pmatrix},
   \label{eq:dynamical_matrix}
\end{equation}
is the dynamical matrix of the chain.
The choice of $\alpha_{A,m}=\alpha_0 - (1+\gamma)-s$ and $\alpha_{B,m}=\alpha_0 - (1-\gamma) - s$ ensures the diagonal elements of $D$ have the same entries $D_{n,n} = 2 + \alpha_0$.
Consequently, after the removal of this diagonal matrix, which shift downward the squared frequency by $\omega^2 = 2 + \alpha_0$, we see that the resulting matrix has chiral or sublattice symmetries (CS/SLS)~\cite{KSUS2019,CXYKT2021,MA2024}.

As a result of the CS/SLS, the bulk squared frequency spectrum
\begin{equation}
    \omega^2_j = (2 + \alpha_0) \pm \sqrt{s^2 + t^2 - 2st\cos\left(\frac{j\pi}{N}\right)},
    \label{eq:bulk_frequency}
\end{equation}
with $j=\pm 1, \pm 2, \ldots \pm N-1$ and $t=\sqrt{1-\gamma^2}$ has two symmetric branches (acoustic and optical) separated by a band gap centered at $\omega^2 = 2 + \alpha_0$.
We consider a chain of $N=33$ sites with $\alpha_0=1$ under OBCs and perform a numerical diagonalization of the eigenvalue problem given in Eq.~\eqref{eq:eigenvalue_problem}.
Figure~\ref{fig:chain_and_spectrum_lineear}(b) shows the dependence of the numerically obtained spectra, $\omega$, against the nonreciprocal strength, $\gamma$, when $s=0.6$.
It shows a discrete spectrum irrespective of $\gamma$, which overall depicts two frequency bands: the acoustic and optical bands located at the lower and upper parts of the spectrum separated by a band gap.

Further, as this dynamical matrix is non-Hermitian, with real frequencies, we can also defined the Hermitian conjugate or left eigenvalue problem $\omega^2 \vec{v} = D^{\dagger}\vec{v}$, whose squared frequencies are given above.
 In this context, $\vec{u}_j$ and $\vec{v}_j$ are the right and left eigenvectors, associated with the real frequency $\omega_j$ and wave number $j$.
These eigenvectors read
\begin{equation}
    u_{j,a,m}\sim d_R^{m-1}, \quad v_{j,a,m} \sim d_L^{m-1},
    \label{eq:skin_modes}
\end{equation}
for the bulk modes, with $d_R = \sqrt{(1+\gamma)/(1-\gamma)}$ and $d_L = \sqrt{(1-\gamma)/(1+\gamma)}$.
As such, the right and left eigenvectors have the same localization ($d_Rd_L = 1$) and are located at the right and left ends of the chain respectively. 
Consequently, these bulk modes are often refereed to as {\it skin modes}.

Furthermore, choosing a chain with an odd number of sites, $N = 2M - 1$, together with the CS/SLS, ensures the presence of a mode with wave number $j = 0$ and frequency $\omega_0^2 = 2 + \alpha_0$.
Figure~\ref{fig:chain_and_spectrum_lineear}(b) confirms the presence of this  mode within the band gap, with frequency $\omega_0 = \sqrt{3}$ in case $\alpha_0 = 1$ [see horizontal black line].
Its eigenvectors have support on the $A$-sublattice,
\begin{equation}
    u_{0,A,m} = r_R^{m-1}, \quad v_{0,A,m} = r_L^{m-1},
    \label{eq:right_left_eigenvector_edge}
\end{equation}
with $r_R = -s/(1 - \gamma)$ and $r_L = -s/(1 + \gamma)$, while the $B$-sublattice has trivial components, i.e., $u_{0,B,m} = 0$ and $v_{0,B,m} = 0$.
We refer to this mode as {\it edge mode}.

In contrast to the skin modes, the localization properties of the right and left eigenvectors of the edge mode vary with the parameters of the system, $\gamma$ and $s$.
Indeed, if $s<1-\gamma$, both the left and right eigenvectors of the edge mode are located at the left end of the chain as $\lvert r_R \rvert <1$ and $\lvert r_L \rvert < 1$.
If instead, $s^2 < 1 - \gamma^2$ and $s > 1-\gamma$, the location of the right eigenvector switch to the right end of the chain, $\lvert r_R \rvert >1$.
At this stage, it is worth looking at their localization properties.
Since, $\lvert r_R \rvert > \lvert r_L\rvert^{-1}$, it follows that the extends of the right eigenvector is larger than its left counterpart.
Finally, in case, $s^2 > 1 - \gamma^2$, we find $\lvert r_R \rvert>1$ and $\lvert r_L \rvert>1$, leading to the location of the right and left eigenvectors at the right end of the chain.
In addition, the localization of the left eigenvector is larger compared to its right counterpart:  $\lvert r_R \rvert < \lvert r_L\rvert^{-1}$.
To summarize, in terms of the location and localization of the finite-frequency edge mode, we identify three regions, in the $(\gamma, s)$ parameter space,
\begin{align}
& \mathrm{(I)}: s+\gamma< 1, \quad \lvert r_{R} \rvert < 1, \lvert r_L\rvert < 1, \nonumber \\
 & \mathrm{(II)}:   \sqrt{s^2+\gamma^2}<  1< s+\gamma,\quad 
   \lvert r_R\rvert > 1, \lvert r_L \rvert< \lvert r_R\rvert^{-1}<1, \nonumber \\
 & \mathrm{(III)}:\sqrt{s^2+\gamma^2}> 1,\quad \lvert r_R \rvert >1,  1>\lvert r_L\rvert > \lvert r_R\rvert^{-1},\label{regions}
\end{align}
which we recapitulate in Fig.~\ref{fig:chain_and_spectrum_lineear}(c).

\section{\label{sec:skin_breathers}Nonreciprocal edge breathers}

\subsection{\label{subsec:numerical_continuation}Characteristics of nonreciprocal edge breathers}
\begin{figure}[!b]
    \centering 
    \includegraphics[width=\textwidth]{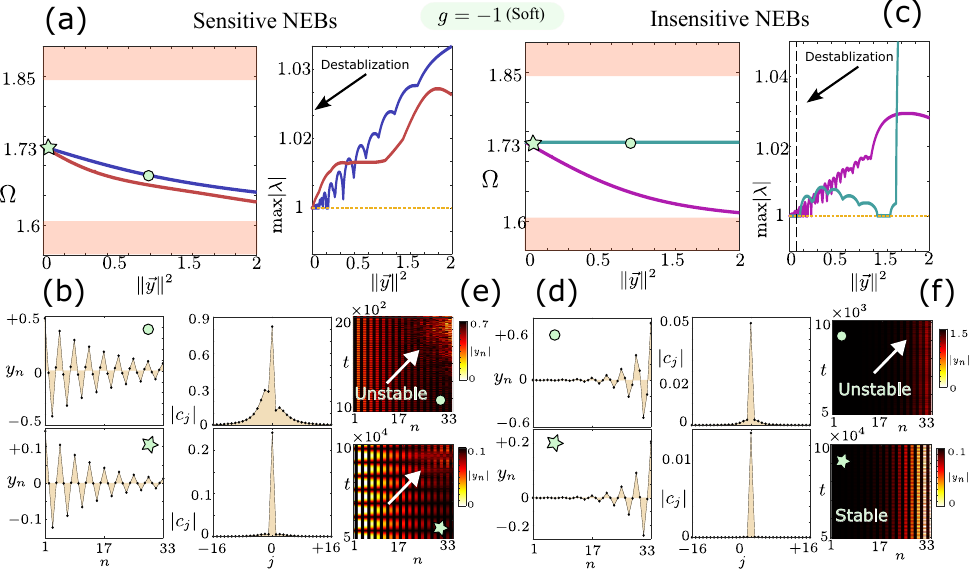}
    \caption{
            Effects of softening nonlinearity ($g = -1$) on families of NEBs emerging from the edge mode with frequency $\omega_0 = 1.73$.
            In all cases, $s = 0.6$, $N = 33$, and $\alpha_0 = 1$.
            (a) Left panel: dependence of the nonlinear frequency $\Omega$ on the amplitude $\lVert \vec{y} \rVert^2$, for $\gamma = 0.3$ (dark blue) and $\gamma = 0.825$ (dark red), corresponding to regions (I) and (III), respectively.
            Right panel: dependence of the magnitude of the largest Floquet eigenvalue, $\max|\lambda|$, on $\lVert \vec{y} \rVert^2$ for the families in (a).
            (b) Profiles of the (left) displacement and (right) normal mode coordinates of representative NEBs with $\gamma = 0.3$ [dark-blue curve in (a)].
            The bottom panels correspond to the NEB with $\lVert \vec{y} \rVert^2 = 0.076$ [green star in (a)], while the top panels correspond to the one with $\lVert \vec{y} \rVert^2 = 1$ [green dot in (a)].
            (c) Same as (a), but for $\gamma = 0.6$ [dark blue, region (II)] and $\gamma = 0$ (dark red).
            (d) Same as (b), but for representative cases belonging to the family of NEBs with (II) $\gamma=0.6$ [cyan curve in (c)].
            (e) Time evolution of representative NEBs with (bottom) $\lVert \vec{y} \rVert^2 = 0.076$ and (top) $\lVert \vec{y} \rVert^2 = 1$ for $\gamma = 0.3$ in region (I).
            The initial conditions for the numerical integration correspond to the profiles shown in (b), perturbed by a deviation vector $\delta y_n = 5 \times 10^{-3} y_n$.
            (f) Same as (e), but for the NEBs shown in (d).
            The arrows indicate dynamical instabilities.    
    }
    \label{fig:fig_continuation_01a}
\end{figure}

We take a small amplitude wave function of the edge mode in the linear limit as the initial guess for a shooting method~\cite{KPGV2009,PVSKG2009}.
The outcome of this process provides an initial state for the family of nonreciprocal edge breathers (NEBs), which we then follow using a nonlinear pseudo-arclength solver~\cite{DKK1991,KPGV2009,PVSKG2009,CK2019}.
Both the shooting method and the solver treat the frequency $\Omega$ and the amplitude $y_n$ as free parameters and the NEBs are obtained with phases such that the $\dot{y}_n=0$.
In addition, we perform the Floquet stability analysis by numerically diagonalizing the monodromy matrix constructed around the obtained nonlinear solution.
The direct numerical integration of the equations of motion and variational equations is carried out using a Runge–Kutta technique of order $8$ based on the Dormand–and-Prince method, denoted $\mathrm{DOP853}$, which ensures high computational accuracy~\cite{HNW1993,DMMS2019,freelyDOP853}.
In what follows, we set $\alpha_0 = 1$.

We first consider the softening case with $g=-1$, fixing the value of $s=0.6$, see dashed line in Fig.~\ref{fig:chain_and_spectrum_lineear}(c).
In the left panel of Fig.~\ref{fig:fig_continuation_01a}(a), we show the dependence of the nonlinear frequency, $\Omega$ against the amplitude, $\lVert \vec{y}\rVert^2$, for representative cases in regions (I) and (III) with $\gamma = 0.3$ (dark-blue) and $\gamma = 0.825$ (dark-red) [$\lVert \cdot \rVert^2$ stands for the Euclidean norm].
These cases are presented in Fig.~\ref{fig:chain_and_spectrum_lineear}(c), by the yellow star and dot symbols.
The figure shows nonlinear frequencies, emerging from $\omega_0 = 1.73$ at small amplitude, and decreasing toward the acoustic band as the amplitude grows.
We illustrate examples of the displacement profiles of these families of NEBs in case (II) $\gamma = 0.3$ for weak and moderate nonlinearities with $\lVert \vec{y} \rVert^2 = 0.076$ [bottom panels of Fig.~\ref{fig:fig_continuation_01a}(b)], and $\lVert \vec{y} \rVert^2 = 1$ [top panels of Fig.~\ref{fig:fig_continuation_01a}(b)] respectively.
Clearly the displacement profiles of these NEBs are localized at the left end of the chain and exhibit displacements on the $B$-sublattice, which tend to grow with increasing amplitude as shown in the left panels of Fig.~\ref{fig:fig_continuation_01a}(b).
This indicates that the presence of onsite nonlinear springs breaks the CS/SLS symmetry of the lattice.

We project the displacement profiles of the NEBs above within the normal mode basis following the bi-orthogonalization, $\vec{y}=\sum_j c_j \vec{u}_j$~\cite{KEBB2018}.
The results of these computations are displayed in the right panels of Fig.~\ref{fig:fig_continuation_01a}(b).
Overall, the coefficients $\lvert c_j\rvert$ grow monotonically with the total amplitude $\lVert \vec{y} \rVert^2$, with the $c_j$ and $y_n$ keeping comparable magnitudes. 
As a result, the nonlinear coupling between the edge mode and the skin modes becomes increasingly pronounced as $\lVert \vec{y} \rVert^2$ grows and the frequency of the NEBs approaches the acoustic band.
Note that due to the softening nonlinearity, bulk modes from the acoustic band contribute most significantly compared to their optical counterparts, as illustrated in the top-right panel of Fig.~\ref{fig:fig_continuation_01a}(b).

\begin{figure}[!t]
    \centering 
    \includegraphics[width=\textwidth]{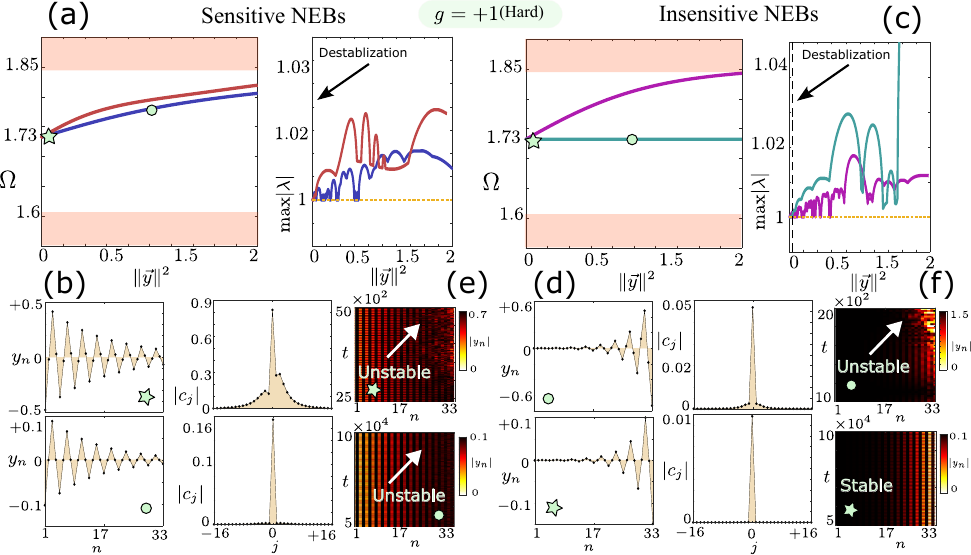}
    \caption{
            Effects of hardening nonlinearity ($g = 1$) on families of NEBs emerging from the edge mode with frequency $\omega_0 = 1.73$.
            In all cases, $s = 0.6$, $N = 33$, and $\alpha_0 = 1$.
            (a) Left panel: dependence of the nonlinear frequency $\Omega$ on the amplitude $\lVert \vec{y} \rVert^2$, for $\gamma = 0.3$ (dark blue) and $\gamma = 0.825$ (dark red), corresponding to regions (I) and (III), respectively.
            Right panel: dependence of the magnitude of the largest Floquet eigenvalue, $\max|\lambda|$, on $\lVert \vec{y} \rVert^2$ for the families in (a).
            (b) Profiles of the (left) displacement and (right) normal mode coordinates of representative NEBs for $\gamma = 0.3$ [dark-blue curve in (a)].
            The bottom panels correspond to a representative NEB with $\lVert \vec{y} \rVert^2 = 0.039$ [green star in (a)], while the top panels correspond to the NEB with $\lVert \vec{y} \rVert^2 = 1$ [green dot in (a)].
            (c) Same as (a), but for $\gamma = 0.6$ [dark blue, region (II)] and $\gamma = 0$ (dark purple).
            (d) Same as (b), but for representative cases belonging to the family of NEBs with (II) $\gamma=0.6$ [cyan curve in (c)].
            (e) Time evolution of the representative NEBs with (bottom) $\lVert \vec{y} \rVert^2 = 0.039$ and (top) $\lVert \vec{y} \rVert^2 = 1$ for $\gamma = 0.3$ in region (I).
            The initial conditions for the numerical integration correspond to the profiles shown in (b), perturbed by a deviation vector $\delta y_n = 5 \times 10^{-3} y_n$.
            (f) Same as (e), but for the NEBs shown in (d).
            The arrows indicate dynamical instabilities.
    }
    \label{fig:fig_continuation_02a}
\end{figure}

We also examine the linear stability of these NEBs~\cite{MA1998,S2001,FG2008,CXYKT2021}.
The numerical calculations using the Floquet analysis, leads to $2\times N$ complex Floquet eigenvalues, $\lambda$. 
It follows that the linear stability of a NEB is guaranteed when the modulus of all Floquet eigenvalues satisfies $\lvert \lambda \rvert = 1$; otherwise, the solution is linearly unstable~\cite{M2025}.
In the right panel of Fig.~\ref{fig:fig_continuation_01a}(a), we present the dependence of the modulus of the largest Floquet eigenvalue against the amplitude.
In both cases, we see that these families of NEBs destabilized almost immediately, with amplitudes of the first destabilization being $\lVert \vec{y} \rVert^2 = 1.3\times 10^{-6}$  and $\lVert \vec{y} \rVert^2 = 4.9\times 10^{-6}$ in cases (I) $\gamma=0.3$ (dark-blue) and (III) $\gamma = 0.825$ (dark-red) respectively.

Let us now look at a representative family of NEBs in region (II)  with $\gamma = 0.6$, see square symbol in Fig.~\ref{fig:chain_and_spectrum_lineear}(c).
Figure~\ref{fig:fig_continuation_01a}(d) shows its $\Omega$–$\lVert \vec{y}\rVert^2$ curve, highlighted in dark cyan.
Surprisingly we find that the nonlinear frequency of the NEBs remains practically constant at the frequency of the linear edge mode, $\Omega = 1.73$, as the $\lVert \vec{y}\rVert^2$ increases.
Consequently, this leads to the emergence of insensitive NEBs.
For comparison, we plot the same numerical results for the reciprocal case with $\gamma=0$ (topological KG chain~\cite{CXYKT2021}), and observe a clear decrease of $\Omega$ with increasing nonlinearity (dark-purple curve). 
It follows that nonlinearity induces significant effects over the interval $\lVert \vec{y}\rVert^2 \in (0,2]$.

The left panels of Fig.~\ref{fig:fig_continuation_01a}(d) show the displacement profiles of representative insensitive NEBs for weak ($\lVert \vec{y} \rVert^2 = 0.076$) and moderate ($\lVert \vec{y} \rVert^2 = 1$) nonlinearities. 
These profiles depict localization at the right end of the chain, with the support on the $B$ sublattice increasing as the nonlinearity grows.
Consequently, this indicates that the mechanisms responsible for the insensitive NEBs do not rely on CS/SLS.s
We express the displacement profiles of these insensitive NEBs, within the normal mode variables, $c_j$.
The resulting $\lvert c_j\rvert$ are shown in the right panels of Fig.~\ref{fig:fig_continuation_01a}(b).
In contrast to regions (I) and (III) [see the right panels of Fig.~\ref{fig:fig_continuation_01a}(b)], the coefficients $\lvert c_j \rvert$ remain consistently small, being at least one order of magnitude lower than the corresponding displacements $y_n$.
Moreover, we find that the contribution from the skin modes is negligible compared to that of the edge mode, $\lvert c_{j\neq  0}\rvert \ll \lvert c_0\rvert$, and tends to mildly increase with growing nonlinearity [see bottom and top panels of Fig.~\ref{fig:fig_continuation_01a}(d)].

In the right-panel of Fig.~\ref{fig:fig_continuation_01a}(c), we show the dependence of the modulus of the largest Floquet eigenvalue $\lvert \lambda \rvert$ against the amplitude, $\lVert \vec{y}\rVert^2$.
We find that the first destabilization of the family of insensitive NEBs occurs at an amplitude $\lVert \vec{y} \rVert^2 = 0.15$, larger than those observed in regions (I) and (III) as well as in the reciprocal case.
It follows that insensitive NEBs, display enhanced linear stability when compared to their sensitive counterparts.
To highlight this enhanced stability, we plot in the bottom panels of Figs.~\ref{fig:fig_continuation_01a}(e) and \ref{fig:fig_continuation_01a}(f) the spatio-temporal evolution of the displacements $y_n(t)$ for representative NEBs at $\lVert \vec{y}\rVert^2 = 0.076$, corresponding to cases (I) $\gamma = 0.3$ and (II) $\gamma = 0.6$, respectively [see the bottom panels of Figs.~\ref{fig:fig_continuation_01a}(b) and (d)].
The initial conditions for our numerical integration are the NEBs above, to which we added a small deviation $\delta y_n \propto 5\times 10^{-3}$.
Consequently, due to instabilities, the NEBs with $\gamma =0.3$ radiates as pointed by the arrow at the bottom panel of Fig.~\ref{fig:fig_continuation_01a}(e).
On the other hand, the dynamics of the insensitive NEBs with $\gamma=0.6$, remains stable till the final time of evolution, $t=10^{5}$ [bottom panel of Fig.~\ref{fig:fig_continuation_01a}(f)].
Moreover, the representative NEBs with $\lVert \vec{y}\rVert^2 = 1$ in regions (I) and (II) [top panels of Figs.~\ref{fig:fig_continuation_01a}(b) and \ref{fig:fig_continuation_01a}(d)] display radiative dynamics together with amplitude growth at the right end of the chain, as evidenced by the colormaps in the top panels of Figs.~\ref{fig:fig_continuation_01a}(e) and \ref{fig:fig_continuation_01a}(f).
These observations are consistent with the stability analysis reported in Fig.~\ref{fig:fig_continuation_01a}(a).

Let us now look at the case with hardening nonlinearity, $g=1$.
Keeping all the numerical  control parameters as above, we focus on the same representative cases in regions (I), (II) and (III) with $\gamma=0.3$, $\gamma=0.6$ and $\gamma=0.825$, respectively the star, square and dot symbols in Fig.~\ref{fig:chain_and_spectrum_lineear}(c). 
In the right panel of Fig.~\ref{fig:fig_continuation_02a}(a), we plot the dependence of the nonlinear frequency, $\Omega$, against the amplitude, $\lVert \vec{y}\rVert^2$, for the families of NEBs at $\gamma=0.3$ (dark-blue) and $\gamma=0.825$ (dark-red) in regions (I) and (III) respectively.
The plot depicts an increasing $\Omega$ for growing $\lVert \vec{y}\rVert^2$, tending toward the optical linear frequency band.
We plot two representative displacement profiles at $\lVert \vec{y} \rVert^2 = 0.039$ and $\lVert \vec{y} \rVert^2 = 1$, belonging to the family of NEBs in region (I) with $\gamma = 0.3$.
These cases, are displayed at the bottom-left and top-left panels of Fig.~\ref{fig:fig_continuation_02a}(b) respectively.
They exhibit localization at the left end of the chain and support also on the $B$ sublattice.
Hence, hardening nonlinearities also result in CS/SLS breaking.

The projections of these displacement profiles onto the normal mode basis [right panels of Fig.~\ref{fig:fig_continuation_02a}(b)] reveal that the coefficients $\lvert c_j \rvert$ have magnitudes comparable to those of the $y_n$ in both cases.
Furthermore, as the nonlinearity increases, the magnitudes of the skin mode variables $\lvert c_{j\neq 0} \rvert$ become comparable to that of the edge mode, $\lvert c_0 \rvert$. 
In particular, contributions from the optical band exceed those from the acoustic band due to the hardening-type nonlinearity [top-right panel of Fig.~\ref{fig:fig_continuation_02a}(b)].
Thus these NEBs are strongly sensitive to nonlinear variations.
Regarding the linear stability of these NEBs, we find that their respective families become unstable quite rapidly as we move away from the linear regime, with the first destabilization occurring at amplitudes $\lVert \vec{y} \rVert^2 = 5\times10^{-6}$ and $\lVert \vec{y} \rVert^2 = 3\times10^{-6}$, as shown in the right panel of Fig.~\ref{fig:fig_continuation_02a}(a).

Turning now to the representative case in region (II) with $\gamma = 0.6$, the right panel of Fig.~\ref{fig:fig_continuation_02a}(c) presents the dependence of $\Omega$ against $\lVert \vec{y}\rVert^2$ (dark-cyan). 
We find a family of insensitive NEBs also for hardening nonlinearities.
Indeed, its $\Omega$ remains practically constant at the frequency of the linear edge mode ($\omega_0=1.73$) as $\lVert \vec{y}\rVert^2$ grows.
To confirm that this amplitude interval ($\lVert \vec{y} \rVert^2 \in (0,2]$) leads to significant nonlinear effects, we superimpose the corresponding curve obtained for the reciprocal lattice with $\gamma = 0$ (dark purple). 
In this case, $\Omega$ increases and nearly enters the optical band as $\lVert \vec{y} \rVert^2 \to 2$.
To confirm that this insensitivity of the NEBs is not due to CS/SLS, in the left panels of Fig.~\ref{fig:fig_continuation_02a}(d), we show the displacement profiles of two representative insensitive NEBs at weak ($\lVert \vec{y} \rVert^2 = 0.039$) and moderate ($\lVert \vec{y} \rVert^2 = 1$) nonlinearities.
We find that they are localized at the right end of the chain, with increasing support on the $B$ sublattice as the nonlinearity grows.
Moreover, the projections of these displacement profiles onto the normal mode basis, $ c_j$, are shown in the right panels of Fig.~\ref{fig:fig_continuation_02a}(d).
The coefficients $\lvert c_j \rvert$ exhibit trends similar to those observed for the softening nonlinearity in region (II).
That is, at both weak and moderate nonlinearities, the coefficients $\lvert c_j \rvert$ remain at least one order of magnitude smaller than their corresponding displacements, $y_n$.
Furthermore, the magnitude of the edge mode variable $\lvert c_0 \rvert$ is incommensurate to those of the skin modes $\lvert c_{j \neq 0} \rvert$.

The linear stability analysis of these insensitive NEBs reveals a first destabilization threshold at $\lVert \vec{y} \rVert^2 = 0.04$ [dashed–dashed line in the right panel of Fig.~\ref{fig:fig_continuation_02a}(c)], which is four orders of magnitude higher than the corresponding thresholds in regions (I) and (III) [right panels of Fig.~\ref{fig:fig_continuation_02a}(b)] and in the reciprocal case ($\lVert \vec{y} \rVert^2 = 1 \times 10^{-6}$) [dark-purple curve in the right panel of Fig.~\ref{fig:fig_continuation_02a}(c)].
Consequently, when examining the dynamical behavior of perturbed NEBs with $\lVert \vec{y} \rVert^2 = 0.039$ in regions (I) with $\gamma = 0.3$ and (II) with $\gamma = 0.6$ we find that the former radiates (see white arrow), while the latter remains dynamically stable [bottom panels of Fig.~\ref{fig:fig_continuation_02a}(e) and Fig.~\ref{fig:fig_continuation_02a}(f)].
In contrast, the dynamics of the representative NEBs of these families at moderate nonlinearities, $\lVert \vec{y}\rVert^2=1$, both exhibit unstable dynamics, as illustrated in the top panels of Figs.~\ref{fig:fig_continuation_02a}(e) and~\ref{fig:fig_continuation_02a}(f) in agreement with their Floquet eigenvalues.

\subsection{\label{subsec:perturbation_theory}Multiple scale analysis and spectral sensitivity diagrams}

\begin{figure}[!h]
    \centering
    \includegraphics[width=0.7\textwidth]{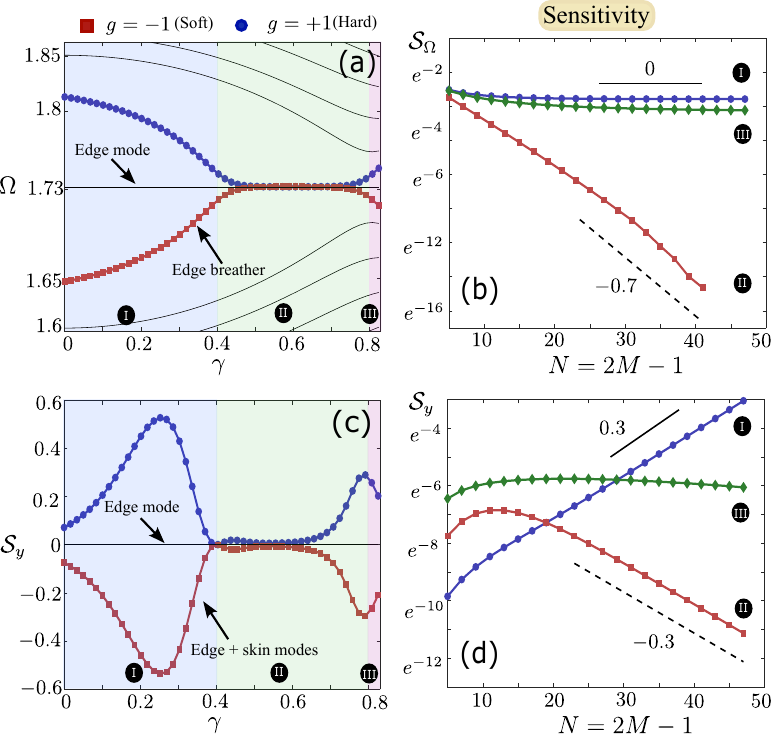}
    \caption{
    Numerically computed spectral sensitivity diagrams of the NEBs.
    (a) Dependence of the nonlinear frequency $\Omega$ of the NEBs on the nonreciprocity parameter $\gamma$.
    The calculations are performed at fixed amplitude $\lVert \vec{y} \rVert^2 = 0.1$ for a chain with $N = 33$, $\alpha_0 = 1$, and $s = 0.6$.
    Blue dot-connected symbols show the numerical results for hardening nonlinearity ($g = 1$), while red square-connected symbols correspond to softening nonlinearity ($g = -1$).
    The black curves represent the variation of the linear spectrum of the lattice, with the horizontal black line indicating the edge mode at frequency $\omega_0 = 1.73$ [see also Fig.~\ref{fig:chain_and_spectrum_lineear}(b)].
    (b) Dependence of the frequency sensitivity factor $\mathcal{S}_\Omega$ (see text for details) on the lattice size $M$ (total number of cells) for representative cases in regions (I) $\gamma = 0.3$, (II) $\gamma = 0.6$, and (III) $\gamma = 0.825$, displayed as blue triangle-, red square-, and green diamond-connected symbols, respectively.
    The straight lines guide the eye toward $\mathcal{S}_\Omega \sim e^{-\beta_\Omega M}$, with $\beta_\Omega = 0.7$ (dashed line) for (II) $\gamma = 0.6$ and $\beta_\Omega = 0$ (solid line) for (I) $\gamma = 0.3$ and (III) $\gamma = 0.825$, obtained from numerical fitting~\cite{SMOOTH}.
    (c) Dependence of the wave-function sensitivity factor $\mathcal{S}_y$ [Eq.~\eqref{eq:wave_function_sensitivity_numerical}] on the nonreciprocity parameter $\gamma$.
    The condition $\mathcal{S}_y = 0$ indicates that the NEB wave function coincides with a nonlinear edge mode.
    (d) Same as (b), but for $\mathcal{S}_y$.
    Numerical fitting yields $\mathcal{S}_y \sim e^{-\beta_y M}$ with $\beta_y = -0.3$ (solid line) for (I) $\gamma = 0.3$ and $\beta_y = -0.3$ (dashed line) for (II) $\gamma = 0.6$.
    }
    \label{fig:frequency_against_nonrecip}
\end{figure}

To explain the insensitivity of the NEBs observed in region (II), we apply a multiple-scale analysis, assuming that their wave function follows~\cite{M2025,NM1979},
\begin{equation}
    y_n (t, T) = \epsilon \left(A(T)u_{0,n} e^{i\omega_0 t} + A^\star (T) u_{0,n} e^{-i\omega_0 t} + \epsilon^2 y_n^{(3)} + \ldots \right),
    \label{eq:ansatz_perturb_01_text}
\end{equation}
with 
$t$ and $T = \epsilon^2 t$ defining the fast and slow temporal variables respectively.
Here $A$ is a slow varying complex amplitude, $\omega_0$ is the linear frequency of the edge mode with wave number $j=0$ and $u_{0,n}$ its associated right eigenvector.
It is worth emphasizing that the terms $\epsilon^2 y_n^{(2)}$ is missing, because of the cubic nonlinearity, see also~\cite{M2025,NM1979}.
Furthermore, the $\{^\star\}$ in Eq.~\eqref{eq:ansatz_perturb_01_text} denotes the complex conjugate.
Substituting Eq.~\eqref{eq:ansatz_perturb_01_text} into Eq.~\eqref{eq:eq_motion} and grouping terms with equal power of $\epsilon$, we find that at the first order $\epsilon$, $\omega_0^2 \vec{u}_0 = D \vec{u}_0$ which corresponds to the linear eigenvalue problem defining the edge mode.

Moving at order $\epsilon^3$, we find that the slow varying complex amplitude obeys a standard nonlinear Schr\"odinger equation
\begin{equation}
    2 i \omega_0 \frac{\partial A}{\partial T}  + 3 \mu \lvert A\rvert^2 A = 0, \quad \mbox{with} \quad  \mu = \frac{ g \sum_{m=1}^{M} (r_L r_R r_R r_R)^{m-1} }{\sum_{m=1}^{M} (r_Lr_R)^{m-1}},
    \label{eq:amplitude_equation_01}
\end{equation}
being its nonlinear coefficient.
Thus Eq.~\eqref{eq:amplitude_equation_01} conserves the norm $\mathcal{A} = \lvert A(T)\rvert^2$ and the energy $\mathcal{H}_A = \mu \lvert A \rvert^{4}/4$.
In particular, its solution reads
\begin{equation}
    A(\epsilon^2 t) = A_0 \exp \left(i\epsilon^2 t \frac{3\mu \lvert A_0 \rvert^2}{2\omega_0}\right),
    \label{eq_app:amplitude_dnls_text}
\end{equation}
where $A_0$ is the amplitude at time $t=0$.
Furthermore, expressing the perturbation parameter $\epsilon$ in terms of the maximum displacement (or amplitude) yields $\lVert \vec{y} \rVert^2 = 4 \epsilon^2 \lvert A_0 \rvert^2 \sum_{m=1}^{M} ( r_R r_L )^{m-1}$, evaluated at phases for which $\dot{y}_n = 0$.
This amplitude, $\lVert \vec{y} \rVert^2$, is a constant parameter and can therefore be used to quantify the nonlinear strength of the NEBs.
It follows that the NEB wave function and its frequency read
\begin{equation}
    y_n (t) = \sqrt{S}\, \left(\frac{u_{0,n}}{\lVert \vec{u}_0\rVert}\right) \cos \left( \Omega t\right),
    \label{eq:nonlinear_waves_normalmodes_text}
\end{equation}
and
\begin{equation}
    \Omega = \omega_0 + \frac{3 g \eta S}{8\omega_0}\left[\frac{(r_L r_R r_R r_R)^{M} - 1}
{\left[(r_R r_R)^{M} - 1\right]\left[(r_L r_R)^{M} - 1\right]} \right],
    \label{eq:nonlinear_frequency_elaborated_text}
\end{equation}
with $S = \lVert \vec{y}\rVert^2$ and $\eta = r_R r_R (r_L r_R - 1)/(r_L r_R r_R r_R - 1)$.
The details of these calculations are provided in Appendix~\ref{app:sec:multiple_scale}.

Equations~\eqref{eq:nonlinear_waves_normalmodes_text} and ~\eqref{eq:nonlinear_frequency_elaborated_text} demonstrate that NEBs emerging from the edge mode exist, provided that a nonlinear correction $\Omega - \omega_0$ to the linear edge mode frequency $\omega_0$.
Moreover, for hardening-type nonlinearities, this correction shifts the NEB's frequency upward, whereas for softening-type nonlinearities, the shift occurs downward.
Interestingly, the magnitude of the nonlinear frequency shift of the NEBs results from the interplay between two distinct contributions: (i) the nonlinear self-interaction of the edge mode, with strength $\propto 3g \lVert \vec{y} \rVert^2 /(8\omega_0)$, which on its own tends to induce a frequency change and (ii) its nonorthogonality, which introduces a competing effect through the imbalance between its left and right eigenvectors ($r_L$ and $r_R$).
Remarkably, this interplay holds for both stationary and oscillating nonreciprocal edge modes~\cite{MA2024}.

To meaningfully analyze the dependence of the nonlinear frequency shift on the system’s control parameters, it is useful to define the frequency sensitivity factor~\cite{M2025,MA2024}
\begin{equation}
    \mathcal{S}_\Omega = \frac{\partial \widetilde{\Omega}}{\partial S}  = \frac{3g\eta}{8\omega_0^2} \frac{(r_L r_R r_R r_R)^{M} - 1}
{\left[(r_R r_R)^{M} - 1\right]\left[(r_L r_R)^{M} - 1\right]},
    \quad S = \lVert \vec{y} \rVert^2,
    \label{eq:frequency_sensitivity}
\end{equation}
with $\widetilde{\Omega} = (\Omega - \omega_0)/\omega_0$.
This quantity measures the rate at which the nonlinear frequency changes with increasing amplitude.
In particular, for finite but sufficiently large lattice sizes ($N,M \gg 1$), the following regimes emerge.
In region (I), characterized by $\lvert r_L \rvert < 1$, $\lvert r_R \rvert < 1$, and $r_L r_R < 1$, we find $\mathcal{S}_\Omega \sim 1$.
Hence, the corresponding NEBs are sensitive to nonlinear perturbations.
On the other hand, in region (II), where $\lvert r_L \rvert < 1$, $\lvert r_R \rvert > 1$, and $r_L r_R < 1$, it follows that $\mathcal{S}_\Omega \sim (r_L r_R)^M = \exp(-\beta_\Omega M)$ with $\beta_\Omega(r_L,r_R) > 0$.
Consequently, the NEBs become insensitive to nonlinear perturbations breaking the CS/SLS.
Finally, in region (III), defined by $\lvert r_L \rvert > 1$, $\lvert r_R \rvert > 1$, and $r_L r_R > 1$, we again obtain $\mathcal{S}_\Omega \sim 1$.
This analysis provides an explanation for the outcomes in Figs.~\ref{fig:fig_continuation_01a}(a) and~\ref{fig:fig_continuation_02a}(a).

Figure~\ref{fig:frequency_against_nonrecip}(a) shows the numerically obtained dependence of the nonlinear frequency of the NEBs, $\Omega$ against the nonreciprocity, $\gamma$, for fixed values of $s = 0.6$ and nonlinearity $\lVert \vec{y} \rVert^2 = 1$.
The case corresponding to a softening nonlinearity ($g = -1$) is represented by red square-connected symbols, while that of a hardening nonlinearity ($g = +1$) is depicted by blue dot-connected symbols.
For comparison, the linear frequencies of the chain are superimposed as black curves, with the linear edge mode indicated by the horizontal black line within the band gap.
This figure shows that when the nonreciprocal strength lies within region (I), i.e., for $\gamma < 0.4$, the nonlinear frequency of the NEBs significantly deviates from its linear counterpart (see the black horizontal line within the gap).
This results in sensitive NEBs, as expected from a topological lattice in which the nonlinearity breaks the CS/SLS (see e.g., Refs.~\cite{CXYKT2021,MS2021}).
As $\gamma \rightarrow 0.4$, marking the transition between regions (I) and (II), the nonlinear frequency of the NEBs gradually decreases.
It follows that within region (II) ($0.4 < \gamma < 0.8$), the nonlinear frequency of the NEBs coincides with that of the linear edge mode, $\Omega \approx \omega_0= 1.73$, corresponding to the regime of insensitive NEBs.
Beyond region (II), when $\gamma \rightarrow 0.8$  and $\gamma > 0.8$ , the nonlinear frequency increases with growing $\gamma$, indicating that the NEBs become sensitive again to nonlinearity.

Figure~\ref{fig:frequency_against_nonrecip}(b) displays the dependence of the frequency sensitivity factor against the lattice size.
We numerically evaluate the $\mathcal{S}_\Omega$ for three representative cases: (I) $\gamma = 0.3$, (II) $\gamma = 0.6$, and (III) $\gamma = 0.825$, with $s = 0.6$.
The amplitude is fixed at $\lVert \vec{y} \rVert^2 = 0.1$, and the rate of change $\Delta \widetilde{\Omega} / \Delta S$ is computed using a two-point finite-difference scheme while varying the lattice size $N$ from $5$ to $47$~\cite{SMOOTH}.
We see that for the representative cases in region (I) ($\gamma = 0.3$, blue triangle-connected symbols) and (III) ($\gamma = 0.825$, green diamond-connected symbols), the fitting of $\mathcal{S}_\Omega  \sim e^{-\beta_\Omega M}$ leads to $\beta_\Omega = 0$ (horizontal solid line).
That is to say, the NEBs are sensitive irrespective of the lattice size. 
In contrast, for the representative case in region (II) ($\gamma = 0.6$, red square-connected symbols), the frequency sensitivity factor exhibits an exponential decay with increasing lattice size, following $\mathcal{S}_\Omega \sim e^{-\beta_\Omega M}$ with $\beta_\Omega = 0.7$, as indicated by the black dashed line. 
It follows that $\Omega=\omega_0$ for sufficiently large lattices, e.g., $M>25$ cells, where the $\mathcal{S}_\Omega$ becomes of the order of the machine precision.

We can further quantify the degree to which the edge mode couples to the surrounding skin modes in presence of nonlinearity.
To this end, we project the NEB wave function [Eq.~\eqref{eq:nonlinear_waves_normalmodes_text}] onto the normal mode variables $c_j$ with $y_n = \sum_j c_j u_{j,n}$.
Within the perturbative regime, we obtain a single-mode solution $y_n(t) = c_0(t) u_{0,n}$, with $c_0(t) = \sqrt{S}\cos(\Omega t)/\lVert \vec{u}_0\rVert$ and $c_{j\neq 0}=0$.
Since the dependence of the nonlinear edge mode $\lvert c_0\rvert$ on the nonlinearity is known, it is natural to assume that the relative coupling strength between the edge mode (with wave number $j=0$) and a skin mode (with wave number $j$) follows a similar scaling, namely $\lvert \mathrm{d}c_j/\mathrm{d}c_0\rvert \sim \lvert c_0\rvert/N$.
Physically, this reflects the fact that as the edge mode is stretched by nonlinearity, its couplings to neighboring modes are elongated at comparable rates.
We thus define the wave-function sensitivity factor as
\begin{equation}
    \mathcal{S}_y = \frac{\partial \lVert \vec{q}\rVert^2}{\partial S} \sim \frac{1}{\left[(r_R r_L)^M - 1\right]}, \qquad S = \lVert \vec{y} \rVert^2,
    \label{eq:wavefunction_sensitivity}
\end{equation}
where $\vec{q} = \vec{c}/c_0$, and $\vec{c} = (c_{-M+1}, \ldots, c_{-1}, c_0, c_1, \ldots, c_{M-1})$.
Following an analysis analogous to that presented above, we find that $\mathcal{S}_y \sim 1$ in regions (I) and (III), whereas in region (II) it decays exponentially as $\mathcal{S}_y \sim \exp(-\beta_{y} M)$ with $\beta_y>0$.

\begin{figure}[!th]
    \centering
    \includegraphics[width=0.7\columnwidth]{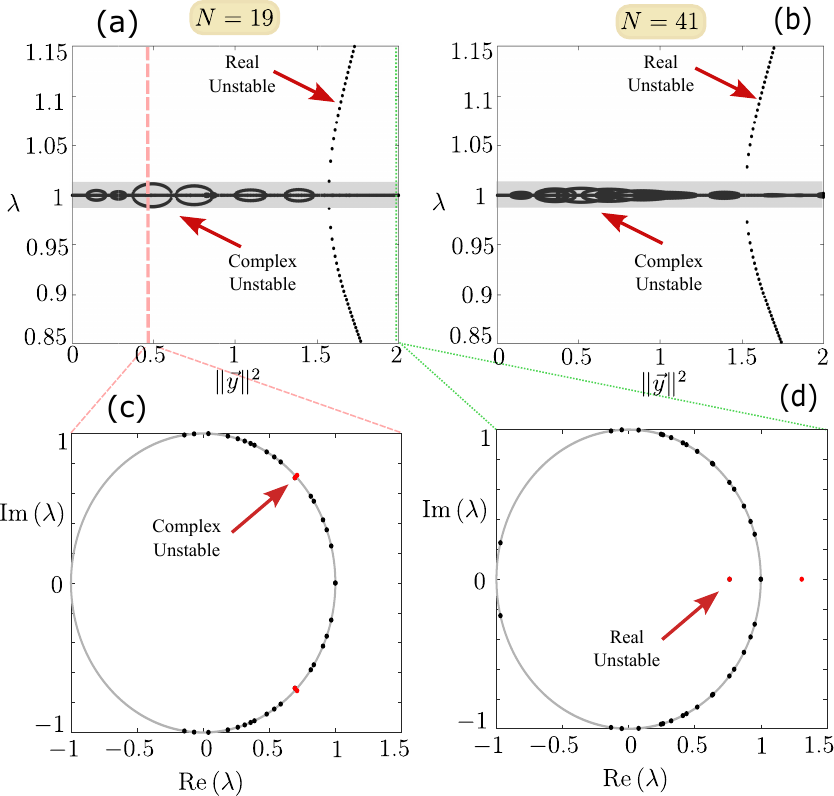}
        \caption{
            Dependence of the modulus of the Floquet eigenvalues, $\lvert \lambda \rvert$, on the amplitude, $\lVert \vec{y}\rVert^2$, for (a) $N=19$ and (b) $N=41$, with $\alpha_0=1$, $g=-1$, $s=0.6$ and $\gamma=0.6$ (II).
            Representative Floquet spectra in the complex plane are shown for (c) $\lVert \vec{y}\rVert^2=0.49$ and (d) $\lVert \vec{y}\rVert^2=2$, in case of $N=19$.
            The red arrows guide the eye to instabilities.
    }
    \label{fig:stability_01}
\end{figure}

Figure~\ref{fig:frequency_against_nonrecip}(c) shows the numerically computed wave-function sensitivity factor~\cite{MA2024}
\begin{equation}
    \mathcal{S}_y = \sum_{j = \pm 1}^{\pm (M-1)} \left\lvert \frac{c_j}{c_0} \right\rvert^2,
    \label{eq:wave_function_sensitivity_numerical}
\end{equation}
obtained from the same continuation data used in Fig.~\ref{fig:frequency_against_nonrecip}(a).
For both softening (dark red) and hardening (dark blue) nonlinearities, the coupling displays a nontrivial variation throughout region (I). 
Starting from $\gamma = 0$, where $\mathcal{S}_y \approx 0.1$, the sensitivity gradually increases, reaching a maximum of $\mathcal{S}_y \approx 0.53$ near $\gamma = 0.25$, before decreasing again as we approach the boundary between regions (I) and (II), i.e. $\gamma \rightarrow 0.4$.
Upon entering region (II) ($\gamma > 0.4$), the sensitivity factor $\mathcal{S}_y$ is strongly suppressed, leading to $\mathcal{S}_y \approx 0$.
As $\gamma$ increases further and approaches $\gamma = 0.8$, marking the boundary between regions (II) and (III), $\mathcal{S}_y$ rises again to nontrivial values with $\max_{\gamma} \mathcal{S}_y \approx 0.3$.

Figure~\ref{fig:frequency_against_nonrecip}(d) shows the dependence of the wave function sensitivity factor $\mathcal{S}_y$ against the lattice size for the same representative cases as in Fig.~\ref{fig:frequency_against_nonrecip}(b).
Wee see that for the representative cases  in regions (I) and (III) with $\gamma = 0.3$ and $\gamma = 0.825$, the $\mathcal{S}_y$ either grows or remains practically constant, following $\mathcal{S}_y \sim e^{-\beta_{y}M}$ with $\beta_y = -0.3$ (bold line) and $\beta_y \approx 0$, respectively.
Indeed, as the lattice size increases, more modes become available, allowing the edge mode to couple with an increasing number of them, which explains why $\mathcal{S}_y$ tends to stay constant or increase.
Surprisingly, for the representative case in region (II) with $\gamma = 0.6$, the wave function sensitivity factor exhibits an exponential decay with growing lattice size, following $\mathcal{S}_y \sim e^{-\beta_y M}$ with $\beta_y = 0.3 $ (black dashed line) in agreement with the analytical results above.

In summary, we demonstrated that the nonreciprocal topological KG of nonlinear classical oscillator supports finite-amplitude edge modes in sufficiently large lattices, despite the absence of CS/SLS.

\subsection{Linear stability of nonreciprocal edge breathers and finite size effects}
It is worth commenting in more detail on the instabilities of the insensitive NEBs at weak and moderate nonlinearities, as shown in Figs.~\ref{fig:fig_continuation_01a}[(a),(c)] and \ref{fig:fig_continuation_02a}[(a),(c)].
Figures~\ref{fig:stability_01}(a) and (b) display the modulus of the Floquet eigenvalues, $\lvert \lambda \rvert$, against the amplitude $\lVert \vec{y} \rVert^2$ for two lattice sizes, $N=19$ [Fig.~\ref{fig:stability_01}(a)] and $N=41$ [Fig.~\ref{fig:stability_01}(b)] with $\gamma = 0.6$, $s=0.6$ and $g = -1$, mapping within region (II) in Fig.~\ref{fig:chain_and_spectrum_lineear}(c).
For both lattice sizes, we observe the emergence of complex instabilities starting at weak amplitudes, $\lVert \vec{y} \rVert^2 \to 0$, and extending to moderate ones up to approximately $\lVert \vec{y} \rVert^2 \simeq 1.6$.
This behavior becomes particularly evident when inspecting the Floquet eigenvalues in the complex plane for a representative insensitive NEB with $\lVert \vec{y} \rVert^2 = 0.49$ in the case $N=19$.
As shown in Fig.~\ref{fig:stability_01}(c), two pairs of Floquet eigenvalues (red dots) lie outside the unit circle (gray circle), unambiguously signaling the presence of complex instabilities.
Interestingly, we find that the onset of these complex instabilities depends on the lattice size since their patterns shape differently in cases of $N=19$ and $N=41$, Figs.~\ref{fig:stability_01}(a)-(b). 
Moreover, the width of the instability band for $N=19$ is significantly wider than that of $N=41$, as highlighted by the gray shaded regions in Figs.~\ref{fig:stability_01}(a) and (b).
This observation indicates that the strength of the complex instabilities decreases with increasing lattice size.
Consequently the instabilities observed in Sec.~\ref{subsec:numerical_continuation} are due to finite size effects.

For amplitudes beyond $\lVert \vec{y}\rVert^2 \approx 1.6$, a new instability branch emerges for both lattice sizes, extending to significantly larger values of $\lvert \lambda \rvert$ as the amplitude grows. 
An inspection of the Floquet spectrum in the complex plane for a representative case with $\lVert \vec{y}\rVert^2 = 2$ reveals that this instability is of real type, characterized by a pair of Floquet eigenvalues departing from the unit circle along the real axis (see red dots).
Furthermore, unlike complex instabilities, this real ones show the same characteristics (pattern) irrespective of the lattice sizes in Fig.~\ref{fig:stability_01}.
This suggests that such instabilities persist even in the large-lattice limit and ultimately determine the stability threshold of the insensitive NEBs.
For completeness, we emphasize that similar results are also observed for hardening-type nonlinearities.

\subsection{\label{subsec:strongly_nonlinear}Strongly nonlinear insensitive nonreciprocal edge breathers}

\begin{figure}[!t]
    \centering
    \includegraphics[width=0.8\textwidth]{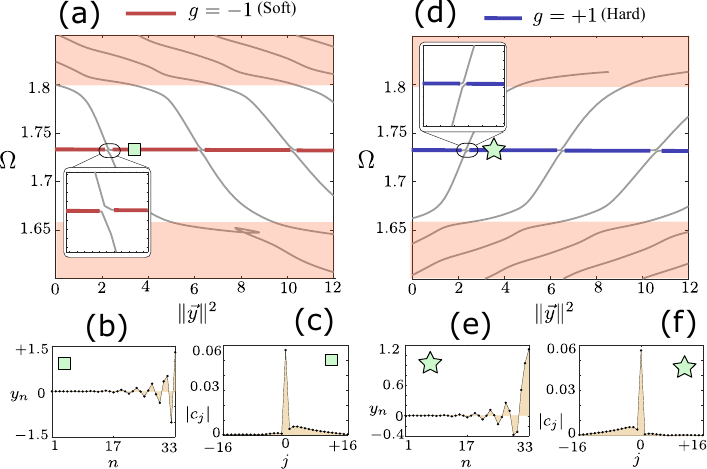}
    \caption{(a) Dependence of the frequency, $\Omega$, against the amplitude $\lVert \vec{y} \rVert^2$ for families of nonlinear breathing modes emerging from all eigenmodes (we show only the cases with $j=0$, $j=1$, $j=2$, $j=3$, $j=4$, $j=5$ and $j=6$) on a chain of $N=33$ sites with $s=0.6$, $\gamma=0.6$, $\alpha_0 = 1$ and $g=-1$.
    (b) Displacement profile of a strongly nonlinear insensitive NEBs, with $\lVert \vec{y} \rVert = 3.4$, and $\Omega = 1.73$.
    (c) Normal mode representation of (b).
    (d) Same as (a)  but for the hardening-type nonlinearity with $g=1$.
    We plot the families of solutions bifurcating form the eigenmodes with $j=-6$, $j=-5$, $j=-4$, $j=-3$, $j=-2$, $j=-1$ and $j=0$.
    (e) Same as (b) but for $g=1$.
    (f) Same as (c), but for (e).
    }
    \label{fig:strong_frequency_against_nonrecip}
\end{figure}

Let us now examine what happens in region (II), within the strong nonlinear regime beyond the validity of the multiple-scale analysis.
We first consider the case of a softening nonlinearity ($g = -1$) with $\gamma = 0.6$ and $s = 0.6$.
We numerically generate using the pseudo-arclength solver, the families of nonlinear nonreciprocal breathing states emerging from all linear modes.
In Fig.~\ref{fig:strong_frequency_against_nonrecip}(a), we plot the dependence of the frequency against the amplitude for the families of solutions  emerging from the modes with $j = 0, +1, +2, +3, +4, +5$, and $+6$ (gray curves).
For completeness, we note that the existence of such nonreciprocal breathing states has been demonstrated theoretically in Appendix~\ref{app:sec:multiple_scale}, see also Ref.~\cite{M2025}.
We observe that the family of insensitive NEBs emerging from the edge mode ($j = 0$) have $\Omega \approx 1.73$ at weak and moderate nonlinearities (see also Sec.~\ref{subsec:numerical_continuation}).
In contrast, the families of nonlinear breathing states emerging from the skin modes of the optical bands exhibit a decreasing $\Omega$ as the $\lVert \vec{y}\rVert^2$ grows.
It follows that, the families emerging from the linear modes with $j = 0$ and $j = +1$ inevitably approach each other, leading to an avoided-crossing bifurcation at $\lVert \vec{y} \rVert^2 \approx 2.3$ (see inset).
This avoided crossing is accompanied by an exchange of properties between the two families.
Indeed, the nonreciprocal breathing mode emerging from the skin mode with $j = +1$ adopts a practically constant nonlinear frequency $\Omega \approx 1.73$ when $\lVert \vec{y} \rVert^2 > 2.3$ leading to {\it strongly nonlinear NEBs}.
On the other hand the family originating from the edge mode ($j = 0$) depicts $\Omega$ values which increases over the same amplitude range.

Figure~\ref{fig:strong_frequency_against_nonrecip}(b) illustrates an instance of the displacement profile of these strongly nonlinear insensitive NEBs with $\lVert \vec{y} \rVert^2 = 3.4$,  after the first avoided crossing bifurcation, see inset of Fig.~\ref{fig:strong_frequency_against_nonrecip}(a).
It clearly exhibits large displacements $y_n$, localization at the right end of the chain like the weak amplitude NEBs and support on both $A$- and $B$-sublattices being of the same order within each unit cell.
Interestingly, although the amplitude $\lVert \vec{y} \rVert^2 = 3.4$ is approximately three times larger than that of the insensitive NEB at $\lVert \vec{y} \rVert^2 = 1$ before the avoided crossing [top-left panel of Fig.~\ref{fig:fig_continuation_02a}(d)], the projections of both profiles onto the normal mode basis yield comparable small values of $\lvert c_j \rvert$ [see the top-right panel of Fig.~\ref{fig:fig_continuation_01a}(d) and Fig.~\ref{fig:strong_frequency_against_nonrecip}(c)].
For instance, $\lvert c_0 \rvert \approx 0.05$ for the case with $\lVert \vec{y}\rVert^2 = 1$ [top-right panel in Fig.~\ref{fig:fig_continuation_02a}(d)], while in case of $\lVert \vec{y}\rVert^2 = 3.4$, we find $\lvert c_0 \rvert \approx 0.06$ [Fig.~\ref{fig:strong_frequency_against_nonrecip}(c)].

Furthermore, a cascade of avoided-crossing bifurcations arises between successive families of nonlinear breathing states emerging from the linear modes with consecutive wave numbers. These occur between $j = 0$ and $j = +1$ at $\lVert \vec{y}\rVert^2 = 2.3$ (as discussed above), $j = +1$ and $j = +2$ at $\lVert \vec{y}\rVert^2 = 6.4$, $j = +2$ and $j = +3$ at $\lVert \vec{y}\rVert^2 = 10.5$, and so on.
In addition, similar behavior is also observed for $g = 1$ (hardening-type nonlinearity), as shown in Figs.~\ref{fig:strong_frequency_against_nonrecip}(d)–(f).
Since this type of nonlinearity leads to an increase of the $\Omega$ of the families of nonlinear breathing modes [Fig.~\ref{fig:strong_frequency_against_nonrecip}(d)], the avoided-crossing bifurcations take place between the families emerging from the edge mode and the skin modes of the acoustic band.

\section{\label{sec:conclusion_breathers}Conclusions}
In this work, we have demonstrated a mechanism for realizing robust finite-amplitude edge waves in nonreciprocal topological mechanical lattices that does not rely on nonlinearities preserving chiral or sublattice symmetries.
Focusing on a nonreciprocal topological Klein–Gordon chain of nonlinear classical oscillators, we established the existence of insensitive nonreciprocal edge breathers (NEBs): spatially localized, time-periodic edge waves that bifurcate from the linear edge mode and persist at finite amplitude.
By combining multiple-scale analysis and numerical continuation methods, we show that insensitive NEBs are characterized by a nonlinear frequency that remains equal to that of the corresponding linear edge mode, irrespective of the nonlinear strength within a region of parameter space.
We further show, at a perturbative level, that this spectral insensitivity originates from a competition between the nonlinear self-interaction of the edge mode and its nonorthogonality induced by nonreciprocal couplings and non-Hermiticity of the system.
As a consequence, the nonlinear frequency shift of insensitive NEBs decays exponentially with increasing lattice size, rendering their spectral properties effectively immune to symmetry-breaking nonlinear perturbations in sufficiently large systems.
Beyond the weakly nonlinear regime, we further demonstrate that insensitive NEBs persist in the strongly nonlinear regime through a sequence of avoided-crossing bifurcations between families of nonlinear breathing modes.
Altogether, our results establish a novel and generic pathway for engineering robust nonlinear topological waves in mechanical metamaterials.
These findings open new directions for the design of robust finite-amplitude edge waves in active mechanical, acoustic, and electrical systems.

\begin{acknowledgments}
B.M.M. acknowledge partial funds from the Israel Science Foundation (ISF) and the EU H2020 ERC StG “NASA” Grant Agreement No. 101077954.
V.A. was supported by the EU H2020 ERC StG “NASA” Grant Agreement No. 101077954.
The authors thank Georgios Theocharis for insightful discussions and Henok T. Moges for careful proofreading of parts of this manuscript.
\end{acknowledgments}

\appendix

\section{\label{app:sec:multiple_scale}Multiple scale analysis}
The equations of motion for the nonreciprocal topological Klein–Gordon chain of nonlinear classical oscillators read
\begin{equation}
    \frac{d^2y_n}{dt^2} = - \sum_{l=1}^{N} D_{n,l} y_l - g y_n^3 ,
    \label{eq_app:eq_motion}
\end{equation}
when expressed in terms of the lattice sites rather than unit cells [cf. Eq.~\eqref{eq:eq_motion}].
Here $n = 1,2,\ldots, N$, with $N$ denoting the lattice size; $y_n$ and $dy_n/dt$ ($t$ is the time) are the displacement and momentum at site $n$, respectively; $D$ is the dynamical matrix defined in Eq.~\eqref{eq:dynamical_matrix}; and $g=\pm1$ corresponds to hardening or softening nonlinearities.
We further impose fixed boundary conditions, $y_0 = y_{N+1} = 0$, with $N$ taken to be odd (see main text for details).

Our goal is to demonstrate the existence of nonlinear normal modes in the lattice defined above that satisfy Eq.~\eqref{eq_app:eq_motion}.
To this end, we employ a multiple-scale expansion and seek solutions of the form~\cite[Chap.~6]{NM1979}.
\begin{equation}
    y_n (t, T) = \epsilon \left(y_n^{(1)} + \epsilon^2 y_n^{(3)} + \ldots \right),
    \label{eq_app:ansatz_perturb_01}
\end{equation}
where $t$ and $T=\epsilon^2 t$ are fast and slow time variables.
It follows that
\begin{equation}
    \frac{d^2}{dt^2} = \left(\frac{\partial }{\partial t} + \epsilon^2 \frac{\partial}{\partial T}\right)^2.
    \label{eq_app:derivs_time_variables}
\end{equation}
Substituting Eq.~\eqref{eq_app:derivs_time_variables} within Eq.~\eqref{eq_app:eq_motion}, leads to
\begin{equation}
    \begin{split}
        \frac{\partial^2}{\partial t^2}\left(\epsilon y_n^{(1)} +  \epsilon^3 y_n^{(3)}\right)  &+ 2\epsilon \frac{\partial^2}{\partial T\partial t}\left(\epsilon y_n^{(1)} + \epsilon^3 y_n^{(3)}\right) = \\ 
        &-\epsilon \sum_{l} D_{n, l} y_l^{(1)} - \epsilon^3 \sum_l D_{n, l} y_l^{(3)} - g \epsilon^3 \left[ y_n^{(1)}  \right]^3,
    \end{split}
\end{equation}
in which we neglected the terms of the order $\epsilon^4$ and higher.
Consequently, at order $\epsilon$ we find 
\begin{equation}
    \frac{\partial^2 y_n^{(1)}}{\partial t^2} =  - \sum_{l=1}^{N} D_{n, l} y_{l}^{(1)}.
    \label{eq_app:first_order_perturb_01}
\end{equation}

Seeking solutions of the form
\begin{equation}
    y_n^{(1)} = A(T)u_ne^{i\omega t} + A^\star(T)u_n e^{-i\omega t},
    \label{eq_app:ansatz_first_order_perturb}
\end{equation}
where $\star$ denotes complex conjugation, it follows that
\begin{equation}
    \omega^2 u_n = \sum_l D_{n,l} u_l, \quad \text{or equivalently} \quad \omega^2 \vec{u} = D\vec{u}.
    \label{eq_app:first_order_perturb_02}
\end{equation}
Consequently, the eigenfrequencies read
\begin{equation}
    \omega_j^2 = (2+\alpha_0) \pm \sqrt{s^2 + t^2 - 2st\cos\left(\frac{j\pi}{N}\right)},
    \quad \text{and} \quad \omega_0^2 = 2 + \alpha_0,
\end{equation}
corresponding to the skin modes with $j=\pm1,\ldots,\pm(M-1)$ and the edge mode with $j=0$, respectively.
The associated eigenvectors $u_{j,n}$ are given in Sec.~\ref{sec:system_skin_and_edge_modes} of the main text.
Thus, at first order in $\epsilon$, our ansatz correctly recovers the linear limit of the chain.
For notational simplicity, we hereafter drop the mode index $j$.

Proceeding at the order $\epsilon^3$, we find
\begin{equation}
        \frac{\partial ^2 y_n^{(3)}}{\partial t^2} + \sum_n D_{n, m} y_n^{(3)} = - 2 \frac{\partial^2 y_n^{(1)}}{\partial T \partial t}  - g \left[ y_n^{(1)}  \right]^3,
        \label{eq_app:second_order_perturb}
\end{equation}
with
\begin{equation}
    \left[ y_n^{(1)}  \right]^3 = 3 \lvert A\rvert^2 A \left( u_n \right)^2 u_n e^{i\omega t} + 3 \lvert A\rvert^2 A^\star \left( u_n \right)^2 u_n e^{-i\omega t}   
    + A^3 \left( u_n \right)^3 e^{3i\omega t} + A^{\star\,3}\left( u_n \right)^3 e^{-3i\omega t}.
    \label{eq_app:nonlinear_term_expansion}
\end{equation}
Substituting Eq.~\eqref{eq_app:nonlinear_term_expansion} into Eq.~\eqref{eq_app:second_order_perturb} and canceling the resonant terms, Eq.~\eqref{eq_app:second_order_perturb} reduces to
\begin{equation}
    - 2i\omega u_n e^{i\omega t} \frac{\partial A}{\partial T} -  3g \lvert A\rvert^2 A \left( u_n \right)^2 u_n e^{i\omega t} = 0.
    \label{eq_app:second_order_currated}
\end{equation}

Next considering the bi-orthogonal nature of the eigenvectors~\cite{AGU2020}, we multiply the left side of Eq.~\eqref{eq_app:second_order_currated} with $\vec{v}^\star e^{-i\omega t}$:
\begin{equation}
    \begin{split}
        \left(v_1e^{-i\omega t},  v_2e^{-i\omega t},   v_3e^{-i\omega t}, \right. & \left. \ldots, v_Ne^{-i\omega t}\right)
        \times \\
        &\left[
            -2i \omega
            \begin{pmatrix}
                u_1 e^{i\omega t} \\ u_2e^{i\omega t} \\ u_3e^{i\omega t} \\ \vdots \\ u_Ne^{i\omega t}
            \end{pmatrix}
            \frac{\partial A}{\partial T} - 3 g \lvert A\rvert^2 A
            \begin{pmatrix}
                \left(u_1\right)^2 u_1e^{i\omega t} \\ \left(u_2\right)^2 u_2e^{i\omega t} \\ \left(u_3\right)^2 u_3e^{i\omega t} \\ \vdots \\ \left(u_N\right)^2 u_Ne^{i\omega t}
            \end{pmatrix}
        \right] = 0.
    \end{split}
\end{equation}
As a result, we obtain the envelope equation for the complex amplitude $A$,
\begin{equation}
    2 i \omega \frac{\partial A}{\partial T} + 3 \mu \lvert A\rvert^2 A = 0, \quad \mbox{with} \quad    \mu = \frac{ g \sum_{n=1}^{N} v_n u_n u_n u_n }{\sum_{n=1}^{N} v_n u_n},
    \label{eq_app:dnls}
\end{equation}
being a nonlinear coefficient.

Thus looking for solutions in the form
\begin{equation}
    A(T) = A_0 e^{-i\nu T},
\end{equation}
we find that
\begin{equation}
    \nu = - \frac{3\mu \lvert A_0\rvert^2}{2\omega}.
\end{equation}
Consequently,
\begin{equation}
    A(T) = A_0 e^{iT \frac{3\mu \lvert A_0 \rvert^2}{2\omega}}
    \label{eq_app:amplitude_dnls}
\end{equation}

Replacing Eq.~\eqref{eq_app:amplitude_dnls} within Eq.~\eqref{eq_app:ansatz_first_order_perturb} and Eq.~\eqref{eq_app:ansatz_perturb_01}, leads to
\begin{equation}
    y_n (t) = 2 \epsilon A_0 u_n \cos \left( \Omega t\right),
    \label{eq_app:nonlinear_modes_01}
\end{equation}
with
\begin{equation}
    \Omega = \omega + \frac{3 \epsilon^2 \mu \lvert A_0 \rvert^2}{2\omega}.
    \label{eq_app:nonlinear_freq_01}
\end{equation}

The remaining task is to identify the small parameter $\epsilon$ in terms of the macroscopic control parameters associated with the nonlinearity.
From Eq.~\eqref{eq_app:ansatz_perturb_01}, we observe that at the phase where the momenta vanish, i.e., $\dot{y}n = 0$, the maximal displacement (or amplitude) satisfies
\begin{equation}
    \lVert \vec{y} \rVert = \epsilon \lvert A_0 \rvert  \lVert \vec{u} \rVert ,
\end{equation}
It then follows that
\begin{equation}
    \lVert \vec{y} \rVert^2 = 4 \epsilon^2 \lvert A_0 \rvert^2 \lVert \vec{u} \rVert^2 = 4 \epsilon^2 \lvert A_0 \rvert^2 \sum_{n=1}^{N} u_n^2 ,
    \label{eq_app:expressing_epsilon}
\end{equation}
is a constant coefficient since $\lvert A(T) \rvert^2 = \lvert A_0 \rvert^2$ is an integral of motion of the envelope equation [Eq.~\eqref{eq_app:dnls}], when retaining only first-order terms in $\epsilon$.
We use $\lVert \vec{y} \rVert^2$ as a measure of the nonlinearity strength.
Replacing Eq.~\eqref{eq_app:expressing_epsilon} into Eqs.~\eqref{eq_app:nonlinear_modes_01} and~\eqref{eq_app:nonlinear_freq_01}, we find
\begin{equation}
    y_n(t) = \sqrt{S} \left( \frac{u_n}{\lVert \vec{u} \rVert} \right) \cos(\Omega t), \quad S = \lVert \vec{y} \rVert^2 ,
    \label{eq_app:nonlinear_modes_02}
\end{equation}
with the nonlinear frequency given by
\begin{equation}
    \Omega = \omega + \frac{3 g \lVert \vec{y} \rVert^2}{8 \omega}
    \left[\frac{\displaystyle \sum_{n=1}^{N} v_n u_n^3} {\displaystyle \left( \sum_{n=1}^{N} u_n^2 \right) \left(\sum_{n=1}^{N} v_n u_n \right)} \right].
    \label{eq_app:nonlinear_freq_02}
\end{equation}
This latter expression, upon substituting the analytical forms of $u_{0,n}$ and $v_{0,n}$ and performing summations, is the one employed in the main text [see Eq.~\eqref{eq:nonlinear_frequency_elaborated_text}].

It is worth noting that Eq.~\eqref{eq_app:nonlinear_freq_02} is general and applies to all families emerging from both the edge mode and the skin modes.
Indeed, while we focus here on the edge mode, the corresponding analysis for skin modes in the nonreciprocal monomer Klein–Gordon model can be found in Ref.~\cite{M2025}.

\section{\label{app:sec:sensitivity_factors}Sensitivity factors of the nonreciprocal edge breathers (NEBs)}
From Eq.~\eqref{eq:frequency_sensitivity}, we analyze the frequency sensitivity factor
\begin{equation}
    \mathcal{S}_\Omega = \frac{\partial \widetilde{\Omega}}{\partial S}\sim \frac{(r_L r_Rr_Rr_R)^{M} - 1}{\left[(r_Rr_R)^{M} - 1\right]\left[(r_L r_R)^{M} - 1\right]},
\end{equation}
where $S = \lVert \vec{y} \rVert^2$ and $\widetilde{\Omega} = (\Omega - \omega_0)/\omega_0$.
In the limit of large lattices ($N \gg 1$), different regimes emerge for the nonreciprocal edge breathers (NEBs).

In region (I), defined by $\lvert r_L \rvert < 1$, $\lvert r_R \rvert < 1$, and $r_L r_R < 1$, we find
\begin{equation}
    \mathcal{S}_\Omega \sim 1,
\end{equation}
indicating that the edge mode remains sensitive to nonlinear perturbations that break chiral or sublattice symmetries (CS/SLS).

In region (II), characterized by $\lvert r_L \rvert < 1$, $\lvert r_R \rvert > 1$, and $r_L r_R < 1$, we obtain
\begin{equation}
    \mathcal{S}_\Omega \sim (r_L r_R)^{M} = \exp \left(-\beta_\Omega M\right), \quad \beta_\Omega(r_L,r_R) > 0,
\end{equation}
signaling the exponential insensitivity of NEBs to the nonlinearity above.

Finally, in region (III), where $\lvert r_L \rvert > 1$, $\lvert r_R \rvert > 1$, and $r_L r_R > 1$, it follows that
\begin{equation}
\mathcal{S}_\Omega \sim \frac{(r_L r_Rr_Rr_R)^{M}}{(r_L r_Rr_Rr_R)^{M}} = 1,
\end{equation}
yielding behavior similar to that found in region (I).

In a similar manner, we compute the wave-function sensitivity as
\begin{equation}
    \mathcal{S}_y = \frac{\partial \lVert \vec{q} \rVert^2}{\partial S},
\end{equation}
where $\vec{q} = \vec{c}/c_0$ is the renormalized coordinate vector in the normal mode variables $c_j$, defined through $y_n = \sum_j c_j u_{j,n}$.
This quantity measures how strongly the edge mode couples to the skin modes as the nonlinearity varies.
In particular, while in the perturbative limit, Eq.~\eqref{eq_app:nonlinear_modes_02} yields
\begin{equation}
    c_0(t) = \left( \frac{\sqrt{S}}{\lVert \vec{u}_0 \rVert} \right)\cos(\Omega t), \quad c_j = 0 ,
    \label{eq_app:nonlinear_edge_mode}
\end{equation}
indicating that only the edge mode is populated at leading order.
It is natural to assume that the rate of change of the relative coupling strength of the skin modes with respect to the edge mode, i.e., $c_j/c_0$, follows the same dependence on the nonlinearity as the nonlinear edge mode itself, Eq.~\eqref{eq_app:nonlinear_edge_mode}.
That is,
\begin{equation}
    \frac{\mathrm{d}}{\mathrm{d}S} \left\lvert \frac{c_j}{c_0} \right\rvert^2 \sim \frac{\mathrm{d}}{\mathrm{d}S} \left( \frac{\lvert c_0 \rvert^2}{N} \right) = \frac{1}{N\left[(r_R r_L)^{M} - 1\right]},
\end{equation}
evaluated at phases of the nonreciprocal edge breathers (NEBs) for which $\dot{c}_j = 0$.
Consequently,
\begin{equation}
    \mathcal{S}_{y} = \sum_{j=\pm 1}^{\pm(M-1)} \left\lvert \frac{c_j}{c_0} \right\rvert^2 \sim \frac{1}{\left[(r_R r_L)^{M} - 1\right]}.
    \label{eq:mode_solution}
\end{equation}

We find that in regions (I) and (III), away from the fully nonreciprocal limit ($r_R \approx 1$),
\begin{equation}
\mathcal{S}_{y} \sim (r_L r_R)^{-M} \sim 1,
\end{equation}
which generally leads to large magnitudes of $c_j$ and strong coupling between the skin modes and the edge mode.
On the other hand, in region (II),
\begin{equation}
\mathcal{S}_{y} \sim (r_L r_R)^{-N} \sim \exp \left(-\beta_{y} M\right),
\qquad \beta(r_R) > 0 .
\end{equation}
In this regime, we therefore expect exponentially small variations of the coefficients $c_j$ and correspondingly weak coupling to the edge mode.
These results are incorporated into the main text.


\bibliography{thebiblio}

\end{document}